\documentstyle[mbf,epsbox,preprint]{ptptex}
\title{
Bi-Local Higgs-Like Fields Based on Non-Commutative Geometry
}

\author{
Shigefumi {\sc Naka},
Shinji {\sc Abe},*
Eizou {\sc Umezawa}
and Tetsu {\sc Matsufuji}
}

\inst{
Department of Physics, College of Science and Technology\\
Nihon University, Tokyo 101-8308, Japan
\\
~*Ibaraki Prefectural University of Health Sciences, Ibaraki 300-0394, Japan
}

\preprintnumber{NUP-A-99-16\\hep-th/9910005}

\recdate{
}
\abst{
The bi-local model of hadrons is studied from the viewpoint of non-commutative geometry formulated so that Higgs-like scalar fields play the role of a bridge, the bi-local fields, connecting different spacetime points. We show that the resultant action for Higgs-like scalar fields has a structure similar to that of the linear sigma model. According to this formalism, we can deduce the dual nature of meson fields as the Nambu-Goldstone bosons associated with chiral symmetry breaking and bound states of quarks.
}

\begin{document}
\maketitle

\section{Introduction}

In the resent understanding, the meson fields are believed to be bound states of quarks ($q$) and anti-quarks ($\bar{q}$) represented by gauge invariant states such as $\langle \bar{q_2}\Phi(C)q_1 \rangle_0$ in the sense of QCD; here, $\Phi(C)=Pe^{i\oint_CA_\mu dx^\mu}$ and $P$ is the path-ordering along the path C that connects $\bar{q_2}$ and $q_1$.\cite{meson} \ The QCD approach, however, requires non-perturbative calculations to analyze the bound states, and at the moment, only numerical approaches based on lattice gauge theory are available to study those states. 

The bi-local field proposed originally by Yukawa {\cite{Yukawa}} has been studied by many authors in the context of relativistic two-body models of quark and anti-quark bound states.\cite{bi-local} \ The standpoint of those approaches is rather simple: In general, $\Phi(C)$ is a complicated functional of the path C; when C is a straight line, however, the field $\langle \bar{q_2}\Phi(C)q_1 \rangle_0$ becomes a bi-local field $B(x,y)$ expected to satisfy

\begin{equation}
 \left[ \frac{1}{2}(\partial_x^2+\partial_y^2) +V(x-y) \right]B(x,y)=0, \label{bi-local}
\end{equation}
where $V(x)$ is an effective potential representing the interaction between $q$ and $\bar{q}$. In such a model, we can calculate various phenomenological features of hadrons. In particular, the covariant oscillator quark model (COQM) that uses $V(x) \propto x^2$ is known to yield fairly good results.\cite{COQM} \ From another point of view, however, the ground states of mesons, the pions, are usually understood as Numbu-Goldstone (N.G.) bosons associated with the chiral symmetry breaking of quark dynamics. It is not obvious that the two interpretations, bound states and N.G. bosons, for pions are compatible.

Recently, a possible way of regarding the Higgs field as a gauge field was proposed by Connes and Lott {\cite{Connes}} from the viewpoint of non-commutative geometry (NCG) and developed by many authors.\cite{many authors} \ In their theory, spacetime is regarded as a product of the usual 4-dimensional spacetime and a discrete two-point space. In those theories, the Higgs fields, which cause the gauge symmetry breaking, play the role of a bridge connecting two kinds of matter fields, the Dirac fields $\bar{\Psi}_R(x)$ and $\Psi_L(x)$, at the same spacetime point $x$ (Fig. 1). 

Then, as an extension of this type of symmetry breaking, it may be possible to introduce Higgs-like fields $\Phi(x,y)$ which connect two kinds of matter fields defined at different spacetime points $x$ and $y$, say $\bar{\Psi}_R(x)$ and $\Psi_L(y)$. From our point of view, the two sheets picture in Fig.~2 describe the proper understanding, and $\Psi_{L}$ and $\Psi_{R}$ are treated as fields in one spacetime. In such a case, we may formulate the bi-local field in Eq.~(1), the bound states of $\bar{q}$ and $q$ fields, as Higgs-like fields, which are related with the chiral symmetry breaking at different spacetime points $x$ and $y$ (Fig.~2). The purpose of this paper is, thus, to study the relationship between the picture of the bound-state and that of the Nambu-Goldstone field for some bi-local fields from the point of view of an extended NCG method.
\footnote{
An early attempt following this line is given in Ref.~{\citen{early attempt}}.
}

The basic formulation is presented in the next section within the framework of a $U(1/1)$ toy model. An attempt to extend this toy model to a more realistic QCD-type model is studied in \S 3.  Section 4 is devoted to summary and discussion.

\begin{figure}[t]
\hspace{10mm}
\begin{minipage}{5.5cm}
   \psbox[width=5cm,height=3cm]{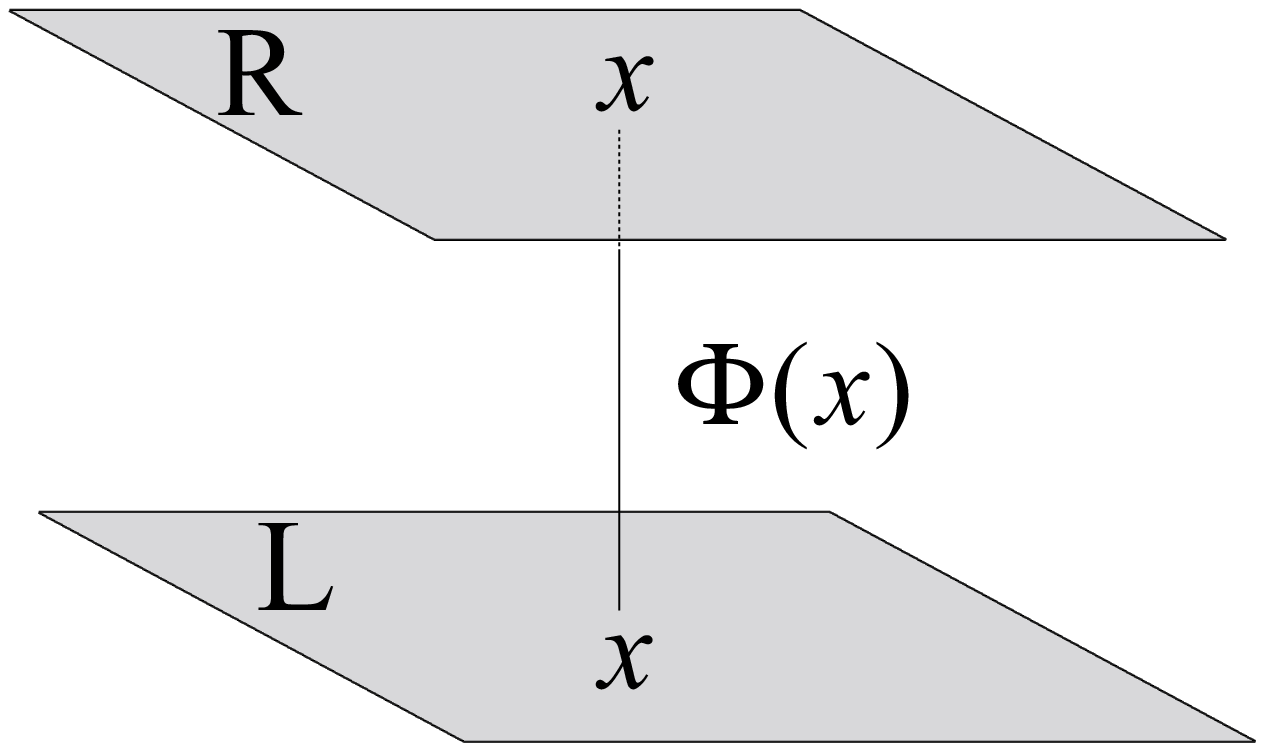}
  \caption{Higgs fields in local NCG.}
\end{minipage}
\hspace{8mm}
\begin{minipage}{5.5cm}
  \psbox[width=5cm,height=3cm]{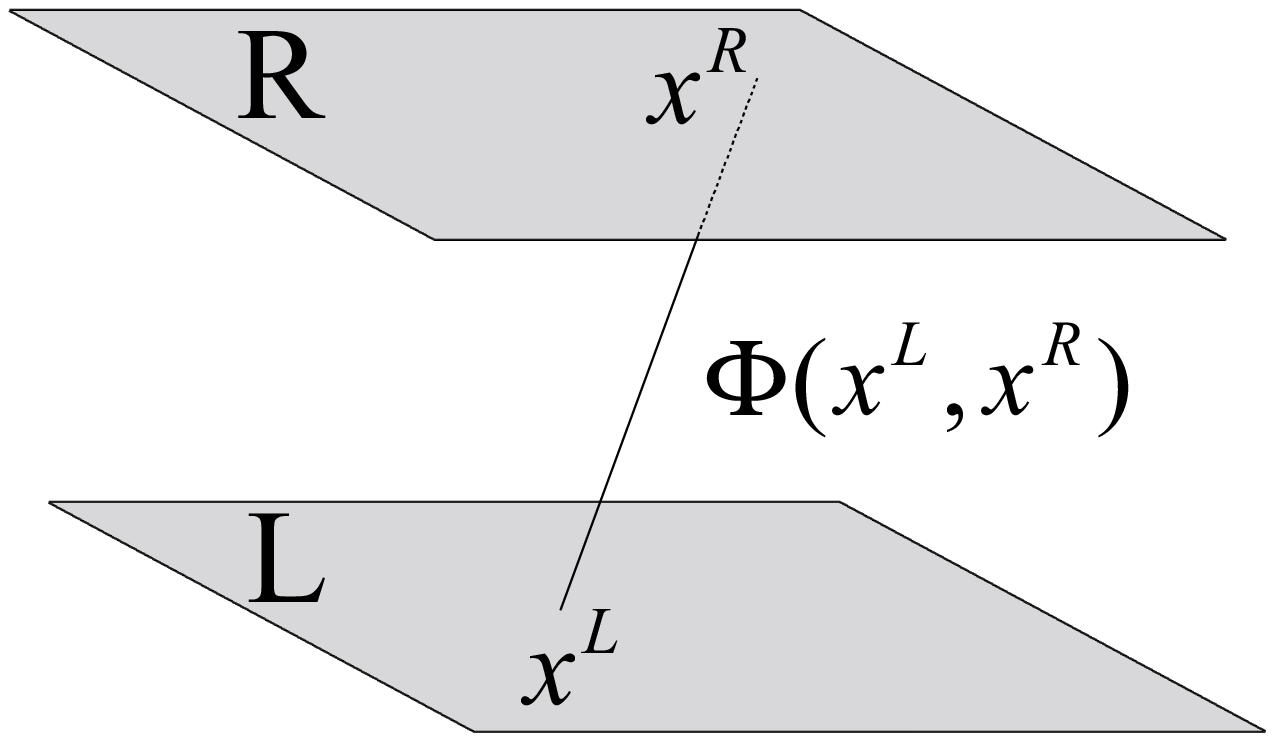}
  \caption{Higgs-like fields in bi-local NCG.}
\end{minipage}
\end{figure}

\section{Formulation of bi-local Higgs-like fields}

The original Connes-Lott formulation for the gauge theory based on NCG is not always convenient for our purposes. In what follows, we use the matrix formulation for such a theory developed by Coquereaux et al. {\cite{matrix formulation}}

To construct the local $U(1/1)$ gauge and Higgs field theories according to the matrix formulation, first, we introduce the off-diagonal matrix $\eta=\frac{1}{2M}\hat{\eta}$, where $\hat{\eta}=\pmatrix{0 & 1 \cr 1 & 0}$, as a kind of coordinate variable of the Dirac field $\Psi(x)=(\Psi_L(x), \Psi_R(x))^T$. Here, $M$ is a parameter with the dimension of mass and the off-diagonal elements "$1$" in $\hat{\eta}$ represent unit matrices in $\Psi_L/\Psi_R$ space.  Second, with analogy to the equation $\partial_{x}f(x)=i[\hat{p},f(x)]$ for ordinary variables, we define the derivative of a $2\times 2$ matrix $A$ with respect to $\eta$ by

\begin{equation}
\partial_\eta A=i[\pi,A\}\equiv i([\pi,A^e]+\{\pi,A^o\}), \label{eta-derivative}
\end{equation}
 \\
where $A^e$ and $A^o$ are the block-diagonal and off-block-diagonal parts of $A$, respectively, and $\pi$ is the momentum operator conjugate to $\eta$ defined by $\{\pi,\eta \}=-{\bf 1}$ in the sense of quantum mechanics. An explicit form of $\pi$ is given by $\pi=-M\hat{\eta}$. Then, one can easily verify the normalization conditions
\begin{equation}
\partial_\eta \eta = -i,~~~~~~\left( {\partial_\eta} \right)^2A=0 \label{eta-condition},
\end{equation}
in addition to the Leibniz rule
\footnote{
The nilpotency $\partial_\eta^2=0$ is satisfied by $\partial_\eta A=\alpha [\pi,A^e]+\beta \{\pi,A^o\}$ for any $\alpha,\beta$. However, the Leibniz rule (\ref{Leibniz}) holds only for $\alpha=\beta$.
}
\begin{eqnarray}
 \partial_\eta (AB) &=& (\partial_\eta A)B+\tilde{A}(\partial_\eta B), \label{Leibniz} \\
 \partial_\eta (A\Psi) &=& (\partial_\eta A)\Psi + \tilde{A}(\partial_{\eta}\Psi),
\end{eqnarray}
where $\tilde{A}=A^e-A^o$. In terms of the matrix $i\pi$, one can also define the operation of $\partial_\eta$ on matter fields by $\partial_\eta \Psi =i\pi \Psi$. 

Now, we can define the covariant derivative operators under the local gauge and discrete transformations. As the generators of $U(1/1)$ , we take the following:

\begin{equation}
T^{L}=\pmatrix{1 & 0 \cr 0 & 0},~~~ T^{R}=\pmatrix{0 & 0 \cr 0 & 1},~~~Q=\pmatrix{0 & 1 \cr 0 & 0},~~~ Q^{\dagger}=\pmatrix{0 & 0 \cr 1 & 0}.\label{U(1/1)-generators}
\end{equation}

\noindent
Then, in the extended spacetime $(x^\mu, \eta)$, one can define the covariant derivatives acting on matrix and matter fields $\Psi$ by
\begin{eqnarray}
 D_\mu &=& \partial_\mu -ig(W^L_\mu T^L + W^R_\mu T^R)=\pmatrix{D^L_\mu & 0 \cr 0 & D^R_\mu}, \label{cov_1}\\
 D_\eta &=& \partial_\eta -ig(\phi Q+Q^\dagger\phi^*)=\partial_\eta -ig\pmatrix{0 & \phi \cr \phi^* & 0}, \label{cov_2} \\
 \tilde{D}_\eta &=& i\pi-ig(\phi Q+Q^\dagger\phi^*)=-ig\pmatrix{0 & \Phi \cr \Phi^* & 0}, \label{cov_3}
\end{eqnarray}
where $D^{L/R}_\mu=\partial_\mu -igW^{L/R}_\mu$ and $\Phi = \phi +\frac{M}{g}$. The last of them one is a covariant derivative operator with respect to $\eta$ acting on matter fields. It should be noted that $[D_\mu, D_\eta ]=[D_\mu, \tilde{D}_\eta ]$ and $D^2_\eta =\tilde{D}^2_\eta -(i\pi)^2$, where the term $(i\pi)^2$ is one origin of symmetry breaking. In terms of these matrices, the Lagrangian density for the $U(1/1)$ gauge theory is given by $\frac{1}{4}{\rm Tr}F^\dagger_{AB}F^{AB} (F_{AB}=\frac{i}{g}[D_A,D_B\},~A,B=\mu,\eta)$. Here $F_{\mu\nu}F^{\mu\nu}$ and $F^\dagger_{\mu\eta}F^{\mu\eta}$ are kinetic terms for gauge and scalar fields, respectively. Furthermore, $F^\dagger_{\eta\eta}F^{\eta\eta}$ gives rise to the potential term for the scalar fields. In addition, the Yukawa coupling between the matter fields and the scalar field is also given by $\bar{\Psi}i\tilde{D}_\eta \Psi$.

Now let us attempt to extend the above gauge-Higgs system to one consisting of local gauge fields and bi-local Higgs-like fields. The matter fields in this case should be written as ${\displaystyle {\mbf \mbf \Psi}=\left[ \matrix{\Psi^L(x^L) \cr \Psi^R(x^R)} \right]}$, where $x^L_\mu$ and $x^R_\mu$ are independent coordinate variables. Accordingly, the indices $L$ and $R$ designate spacetime points in addition to gauge or chiral components in the extended formalism. As for the discrete variables $\eta$ and $\pi$, we try the same form as in the local NCG, but $M$ may be a c-number function of $(x^L-x^R)^2$ if we consider translational invariance. Then, writing $\partial^{L/R}_\mu=\partial/\partial x^{L/R,\mu}$, one can define the following as covariant-derivative operators acting on extended matrix or matter fields:
\begin{eqnarray}
 {\mbf D}^{L\pm}_\mu &=& \pmatrix{ D^L_\mu \pm \partial^R_\mu & 0 \cr 0 & 0 }=(D^L_\mu \pm \partial^R_\mu)T^L,  \label{bi-covariant_1} \\
 {\mbf D}^{R\pm}_\mu &=& \pmatrix{ 0 & 0 \cr 0 & D^R_\mu \pm \partial^L_\mu }=(D^R_\mu \pm \partial^L_\mu)T^R  \label{bi-covariant_2}, \\
 {\mbf D}_\eta &=& \partial_\eta -ig\pmatrix{ 0 & \phi_{LR} \cr \phi_{RL} & 0}, \label{bi-covariant_3} \\
 \tilde{\mbf D}_\eta &=& -ig\pmatrix{ 0 & \Phi_{LR} \cr \Phi_{RL} & 0 }, ~~~~~~\left( \Phi_{LR}=\phi_{LR}+\frac{M}{g} \right), \label{bi-covariant_4}
\end{eqnarray}
where $D^{L/R}_\mu=\partial^{L/R}_\mu-igW^{L/R}_\mu(x^{L/R})$, $\Phi_{LR}=\Phi(x^L,x^R)$, and $\Phi_{RL}=\Phi^\dagger(x^R,x^L)$. Indeed, since the gauge transformations ${\mbf \mbf \Psi}\rightarrow {\cal U}{\mbf \Psi}$ are induced by
\begin{equation}
 {\cal U}= \left[ \matrix{ {\cal U}^L(x^L) & 0 \cr 0 & {\cal U}^R(x^R) } \right] , ~~~~~~~({\cal U}^{L\dagger}{\cal U}^L={\cal U}^{R\dagger} {\cal U}^R={\bf 1}),
\end{equation}
the operators in (\ref{bi-covariant_1}) - (\ref{bi-covariant_4}) undergo the transformations ${\cal U}{\mbf D}^{L/R\pm}_\mu {\cal U}^\dagger ={\mbf D}^{'L/R\pm}_\mu$, ${\cal U}{\mbf D}_\eta {\cal U}^\dagger ={\mbf D}'_\eta$, and so on. Here, the primes indicate covariant-derivative operators, in which the gauge and scalar fields are replaced, respectively, by
\begin{eqnarray}
 W^{L/R}_\mu  \rightarrow  W^{'L/R}_\mu&=&{\cal U}^{L/R}W^{L/R}_\mu {\cal U}^{L/R\dagger}+\frac{i}{g}\,{\cal U}^{L/R}\partial^{L/R}_\mu {\cal U}^{L/R\dagger} \label{gauge transformation_1},  \\
 \Phi_{LR}  \rightarrow  \Phi_{LR}^{'}&=&{\cal U}^L\Phi_{LR}{\cal U}^{R\dagger} . \label{gauge transformation_2}
\end{eqnarray}
In this sense, (\ref{bi-covariant_1}) - (\ref{bi-covariant_4}) define covariant-derivative operators provided (\ref{gauge transformation_1}) and (\ref{gauge transformation_2}) hold. It should be noted, however, that $\Phi_{LR}$ obeys the bi-linear transformation (\ref{gauge transformation_2}). Then, $\phi_{LR}$ does not obey simple bi-liniear transformations under gauge transformations, due to the factor $M$, and vice versa. For the time-being in this section, we assume the transformation (\ref{gauge transformation_2}).

For later purposes, let us introduce the center of mass variables $X_\mu=\frac{1}{2}(x^L_\mu+x^R_\mu)$ and the relative variables $\bar{x}_\mu=x^L_\mu-x^R_\mu$, in terms of which we can write
\begin{eqnarray}
 x^L_\mu &=& X_\mu+\frac{1}{2}\bar{x}_\mu,~~~~~x^R_\mu=X_\mu-\frac{1}{2}\bar{x}_\mu ~, \\
 \partial_\mu &=& \frac{\partial}{\partial X^\mu}=\partial^L_\mu+\partial^R_\mu,~~~~~\bar{\partial}_\mu=\frac{\partial}{\partial \bar{x}^\mu}=\frac{1}{2}\left(\partial^L_\mu-\partial^R_\mu \right).
\end{eqnarray}
It is, thus, obvious that by taking the limit $x^L_\mu,x^R_\mu \rightarrow x_\mu$, the extended derivative operators ${\mbf D}^{L/R+}_\mu$ and ${\mbf D}_\eta$ will tend to the local operators $(\partial_\mu-igW^{L/R})T^{L/R}$ and $D_\eta$, respectively. On the other hand, ${\mbf D}^{L/R -}_\mu$ will tend to $-igW^{L/R}_\mu T^{L/R}$.

The candidates for the field strength constructed out of these extended covariant-derivative operators are
\begin{eqnarray}
 [ {\mbf D}^{L\pm}_\mu,{\mbf D}^{L\pm}_\nu ] &=& [ {\mbf D}^{L\mp}_\mu,{\mbf D}^{L\pm}_\nu ]=-ig\partial_{[\mu}W^L_{\nu]}T^L=-ig{\mbf F}^L_{\mu\nu}, \\ 
 ~[ {\mbf D}^{R\pm}_\mu,{\mbf D}^{R\pm}_\nu ] &=& [ {\mbf D}^{R\mp}_\mu,{\mbf D}^{R\pm}_\nu ]=-ig\partial_{[\mu}W^R_{\nu]}T^R=-ig{\mbf F}^R_{\mu\nu}, \\
 \{ {\mbf D}_\eta,{\mbf D}_\eta \} &=& -2g^2\left(\Phi_{LR}\Phi_{RL}-\frac{1}{g^2}M^2\right)=-ig{\mbf F}_{\eta\eta}.
\end{eqnarray}
In contrast to the case of local covariant-derivative operators, however, $[ {\mbf D}^{L+}_\mu,{\mbf D}_\eta ]$, etc., remain as operators in the following sense:
\begin{eqnarray}
 [{\mbf D}^{L+}_\mu,{\mbf D}_\eta] &=& -ig\left[(\partial_\mu -igW^L_\mu)\Phi_{LR}Q-\Phi_{RL}(\partial_\mu -igW^L_\mu)Q^\dagger \right] , \\
 ~[{\mbf D}^{L-}_\mu,{\mbf D}_\eta] &=& -ig\left[(2\bar{\partial}_\mu -igW^L_\mu)\Phi_{LR}Q-\Phi_{RL}(2\bar{\partial}_\mu -igW^L_\mu)Q^\dagger \right] , \\
 ~[{\mbf D}^{R+}_\mu,{\mbf D}_\eta] &=& -ig\left[-\Phi_{LR}(\partial_\mu -igW^R_\mu)Q+(\partial_\mu -igW^R_\mu)\Phi_{RL}Q^\dagger \right] , \\
 ~[{\mbf D}^{R-}_\mu,{\mbf D}_\eta] &=& -ig\left[ \Phi_{LR}(2\bar{\partial}_\mu +igW^R_\mu)Q-(2\bar{\partial}_\mu +igW^R_\mu)\Phi_{RL}Q^\dagger \right] .~~~~~~~~~~
\end{eqnarray}
In order to obtain the c-number field strength, we have to take the combinations such that
\begin{eqnarray}
 [{\mbf D}^{L+}_\mu + {\mbf D}^{R+}_\mu ,{\mbf D}_\eta] &=& -ig\left[\left\{(\partial_\mu\Phi_{LR})-ig(W^L_\mu\Phi_{LR}-\Phi_{LR}W^R_\mu)\right\}Q \right. \nonumber \\
 & & + \left .\left\{(\partial_\mu\Phi_{RL})-ig(W^R_\mu\Phi_{RL}-\Phi_{RL}W^L_\mu)\right\}Q^\dagger \right]  \nonumber \\
 &=& -ig{\mbf F}^{+}_{\mu\eta} , \label{mu-eta+}\\
 ~[{\mbf D}^{L-}_\mu - {\mbf D}^{R-}_\mu ,{\mbf D}_\eta] &=& -ig\left[ \left\{2(\bar{\partial}_\mu\Phi_{LR})-ig(W^L_\mu\Phi_{LR}+\Phi_{LR}W^R_\mu)\right\}Q \right. \nonumber \\
 & & + \left. \left\{2(\bar{\partial}_\mu\Phi_{RL})+ig(W^R_\mu\Phi_{RL}+\Phi_{RL}W^L_\mu)\right\}Q^\dagger \right] \nonumber \\
 &=& -ig{\mbf F}^{-}_{\mu\eta}. \label{mu-eta}
\end{eqnarray}
Here, ${\mbf F}^{L}_{\mu\nu}$ and ${\mbf F}^{R}_{\mu\nu}$ are local field strengths, while ${\mbf F}_{\eta\eta}$ and ${\mbf F}^{\pm}_{\mu\eta}$ are bi-local field strengths. Therefore, in this model,
\begin{eqnarray}
S_{GH} &=& -\frac{1}{4}tr \int d^4x \left({\mbf F}^L_{\mu\nu}{\mbf F}^{L\mu\nu}+{\mbf F}^R_{\mu\nu}{\mbf F}^{R\mu\nu} \right) \nonumber \\
& & -\frac{1}{2} tr\int d^4x^L \int d^4x^R \left\{({\mbf F}^+_{\mu\eta})^\dagger{\mbf F}^{+\mu\eta}+({\mbf F}^-_{\mu\eta})^\dagger{\mbf F}^{-\mu\eta} + \frac{1}{2}{\mbf F}^\dagger_{\eta\eta}{\mbf F}_{\eta\eta} \right\}~~~~~~ \label{action GH}
\end{eqnarray}
is a natural form of the action defined from these field strengths. This form preserves the $L \leftrightarrow R$ invariance and tends to the action for the local gauge and Higgs fields in the limit $x^L\rightarrow x^R$. The interaction between the matter fields and gauge-Higgs fields should also be given by
\begin{equation}
 S_M=\int d^4x^L\int d^4x^R \overline{\Psi}\left\{ \delta^{(4)}(x^L-x^R)i\gamma^\mu D_\mu+i\kappa\tilde{D}_\eta \right\}\Psi , \label{action M}
\end{equation}
where $\kappa$ is a parameter. Then in terms of component fields, the action $S=S_{GH}+S_M$ can be written as
\begin{eqnarray}
 S &=& -\frac{1}{4}\int d^4x \left( F_{\mu\nu}^LF^{\mu\nu L}+F_{\mu\nu}^RF^{\mu\nu R}\right)  \nonumber \\
 & & + \int d^4x^L\int d^4x^R \left\{ \left|{\cal D}_\mu \Phi_{LR}\right|^2 + 4\left|\bar{\cal D}_\mu \Phi_{LR} \right|^2 \right\} \nonumber \\
 & & -2g^2 \int d^4x^L\int d^4x^R \left( \Phi_{LR}\Phi_{RL}-\frac{1}{g^2}M^2 \right)^2 \nonumber \\
 & & +\int d^4x\overline{\Psi}i\gamma^\mu\left\{ \partial_\mu -ig\left(W^L_\mu T^L+W^R_\mu T^R \right) \right\} \Psi \nonumber \\
 & & + g\kappa \int d^4x^L\int d^4x^R \left(\overline{\Psi}^L\Phi_{LR}\Psi^R+\overline{\Psi}^R \Phi_{RL}\Psi^L \right), \label{action}
\end{eqnarray}
where
\begin{eqnarray}
 {\cal D}_\mu \Phi_{LR} &=& \partial_\mu\Phi_{LR}-ig(W_\mu^L\Phi_{LR}-\Phi_{LR}W_\mu^R), \\
 \bar{\cal D}_\mu \Phi_{LR} &=& \bar{\partial}_\mu\Phi_{LR}-\frac{1}{2}ig(W_\mu^L\Phi_{LR}+\Phi_{LR}W_\mu^R), \label{D_Phi_LR}
\end{eqnarray}
and the contractions ${\cal D}_\mu\Phi_{LR}{\cal D}^\mu\Phi_{RL}$, etc., have been abbreviated as $|{\cal D}_\mu\Phi_{LR}|^2$, and so on. 

In the above action, the potential for $\Phi_{LR}$ has a minimum at $\Phi_{LR}\Phi_{RL}=\frac{1}{g^2}M^2$. There are several ways to choose the vacuum, and the following may coincide in the local limit $x^L \rightarrow x^R$:
\begin{eqnarray}
{\rm ~i}) & ~~ & \Phi_{LR}=\frac{1}{g}M+U_L\tilde{\Phi}_{LR}U_R^\dagger ,  \label{vacuum_1} \\
{\rm ii}) & ~~ &  \Phi_{LR}=U_L(\frac{1}{g}M+\tilde{\Phi}_{LR})U_R^\dagger .  ~~~~~~~~~~~~~~~~~~~~~~~~~~~~~~~~~~~~~~~  \label{vacuum_2}
\end{eqnarray}
Here, $U^{L}$ and $U^{R}$ are unitary-matrix fields that transform as $U^{L/R} \rightarrow {\cal U}^{L/R}U^{L/R}$ under the gauge transformation. $\tilde{\Phi}_{LR}$ is assumed to be a gauge-invariant field. The case (i) is the original configuration of $\Phi_{LR}$ given in the covariant derivative (\ref{bi-covariant_4}), as can be seen by identifying $\phi_{LR}$ with $U_L\tilde{\Phi}_{LR}U_R^\dagger$. Since $\phi_{LR}$ cannot obey the same bilinear transformations as $\Phi_{LR}$, the form of (\ref{bi-covariant_4}) breaks local gauge invariance from the outset, except in the case that $M(\bar{x}) \propto  \delta^4(\bar{x})$. The case (ii) appears to preserve local-gauge invariance. However, the form is not compatible with the original configuration of $\Phi_{LR}$ in (\ref{bi-covariant_4}). We will discuss a modified version of the case (ii) in the next section within the framework of a more realistic model.

Let us now consider a QCD-type gauge theory in the case (i). Then, we need not distinguish $L/R$ gauge structures, and we may put $W^{L/R}_\mu=W_\mu(x^{L/R})$. In this case, the covariant derivatives and the potential term factors in (\ref{action}) can be rewritten in the form
\begin{eqnarray}
 {\cal D}_\mu \Phi_{LR} \! &=& \! U_L \! \left[\partial_\mu \left\{ \frac{1}{g}MU_{RL}+\tilde{\Phi}_{LR} \right\} \! - \! ig(W^{\prime L}_\mu - W^{\prime R}_\mu)\left\{\frac{1}{g}MU_{RL}+\tilde{\Phi}_{LR}\right\} \right] \! U^\dagger_R ,  \nonumber \\
 & & \label{case 1-1}  \\
 \bar{\cal D}_\mu \Phi_{LR} \! &=& \! U_L \! \left[\bar{\partial}_\mu \left\{ \frac{1}{g}MU_{RL} + \tilde{\Phi}_{LR} \right\} \! - \! \frac{i}{2}g(W^{\prime L}_\mu + W^{\prime R}_\mu)\left\{\frac{1}{g}MU_{RL}+\tilde{\Phi}_{LR}\right\}  \right] \! U^\dagger_R , \nonumber \\
 & & \label{case 1-2}  
\end{eqnarray}
and
\begin{equation}
 \Phi_{LR}\Phi_{RL} - \frac{1}{g^2}M^2 = \frac{1}{g}M(U_{LR}\tilde{\Phi}_{LR}+U_{RL}\tilde{\Phi}_{RL})+\tilde{\Phi}_{LR}\tilde{\Phi}_{RL}, \label{case 1-3}
\end{equation}
where we have used the abbreviations $U_{LR}$ for $U_LU^\dagger_R$, $U_{RL}$ for $U_RU^\dagger_L$, and
\begin{equation}
 W_\mu^{\prime K} ~~\mbox{for}~~ W^{K}_\mu +\frac{i}{g}U^\dagger_{K}\partial_\mu U_{K},~~~(K=L,R). \label{W prime-0}
\end{equation}

From the expression (\ref{case 1-3}), it is obvious that linear terms of $M$ give rise to mass terms for the $\tilde{\Phi}_{LR}$ field. If we decompose the $\tilde{\Phi}_{LR}$ field into its real components in such a way that
\begin{equation}
 \tilde{\Phi}_{LR}=\sigma_{LR}+i\pi_{LR},~~~ \tilde{\Phi}_{RL}=\sigma_{RL}-i\pi_{RL},
\end{equation}
then one can say that $\sigma_{LR}$ and $\pi_{LR}$ become massive fields, respectively, for $L/R$ symmetric and $L/R$ anti-symmetric configurations.

As for $U_L$ and $U_R$, it is also interesting to consider
\begin{eqnarray}
 U_L &=& P\exp\left\{ig\int_{(C_L)}^{x^L}dz^\mu W_\mu^L(z) \right\}, \\
 U_R &=& P\exp\left\{ig\int_{(C_R)}^{x^R}dz^\mu W_\mu^R(z) \right\}. 
\end{eqnarray}

\begin{figure}[t]
\hspace{10mm}
\begin{minipage}{5.5cm}
  \psbox[width=5cm,height=3cm]{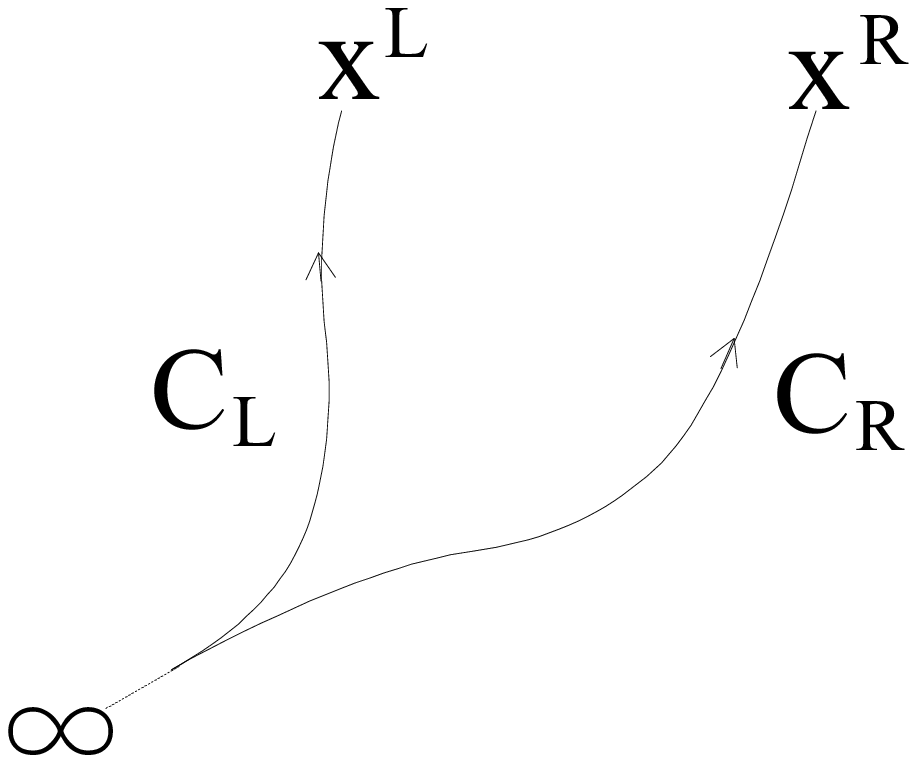}
  \caption{Wilson string along $C_L$ and $C_R$.}
\end{minipage}
\hspace{8mm}
\begin{minipage}{5.5cm}
  \psbox[width=5cm,height=3cm]{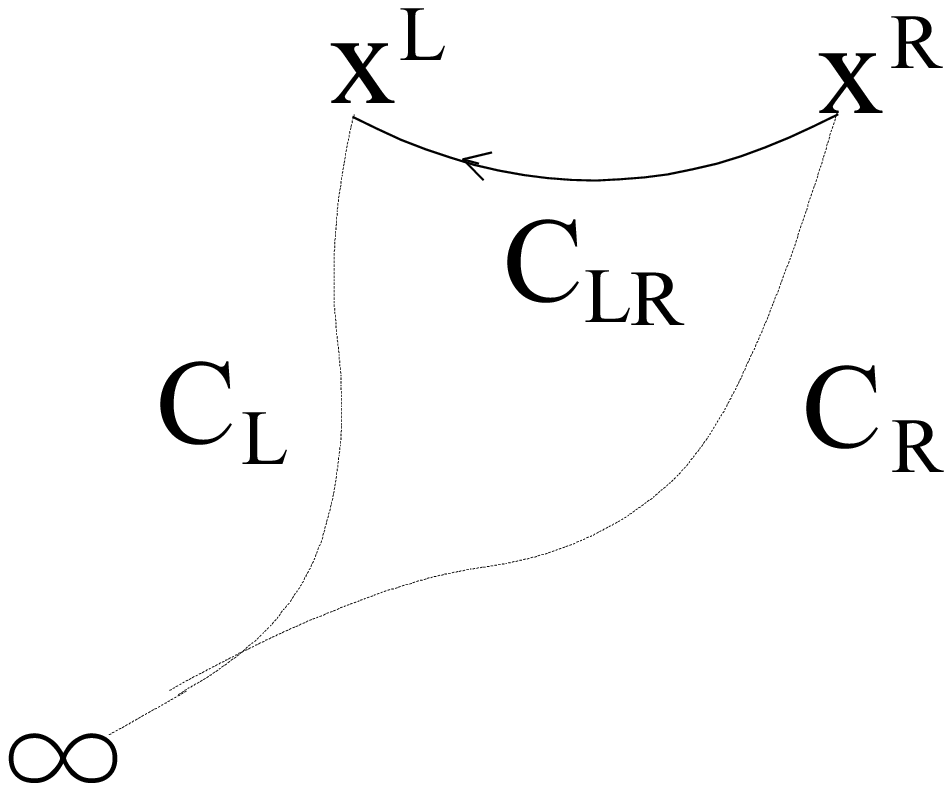}
  \caption{Wilson string along $C_{LR}$.}
\end{minipage}
\end{figure}
\noindent
Here $C_L$ and $C_R$ are paths from infinitely distant places to $x_L$ and $x_R$, respectively, (Fig.~3); according to these choices of $U_L$ and $U_R$, $U_{LR}$ can be understood as
\begin{equation}
 U_{LR}=P\exp\left\{ig\int_{C_{LR}}dz^\mu W_\mu(z) \right\}, \label{Wilson string}
\end{equation}
where $C_{LR}$ is a finite path from $x^R$ to $x^L$ (Fig.~4). If it is necessary, one may define $C_{LR}$ as a straight line, and then, it is possible for $\tilde{\Phi}_{LR}$ to retain its meaning as a bi-local field.

A particularly interesting point regarding the above choice of $U_{L}$ and $U_{R}$ is the fact that $W^{\prime L}_\mu$ and $W^{\prime R}_\mu$ defined in (\ref{W prime-0}) vanish, and thus the action becomes very simple. If we adopt (\ref{Wilson string}) to express the gauge degrees of freedom $U_{L}$ and $U_{R}$  of $\phi_{LR}$, then the action (\ref{action}) is reduced to
\begin{eqnarray}
 S &=& -\frac{1}{2}\int d^4x F_{\mu\nu} F^{\mu\nu}  \nonumber \\
 & & + \int d^4x^L\int d^4x^R \left\{ \left|\partial_\mu \left( \frac{1}{g}MU_{LR}+\tilde{\Phi}_{LR} \right) \right|^2 + 4\left|\bar{\partial}_\mu\left(\frac{1}{g}MU_{LR}+\tilde{\Phi}_{LR}\right) \right|^2 \right\} ~~~~~~~~~~ \nonumber \\
 & & -2g^2 \int d^4x^L\int d^4x^R \left\{ \frac{1}{g}M\left(U_{LR}\tilde{\Phi}_{LR}+U_{RL}\tilde{\Phi}_{RL}\right)+\tilde{\Phi}_{LR}\tilde{\Phi}_{RL}) \right\}^2 \nonumber \\
 & & +\int d^4x\overline{\Psi}i\gamma^\mu \left( \partial_\mu -igW_\mu \right) \Psi \nonumber \\
 & & +g\kappa \int \! d^4x^L \! \int \! d^4x^R \! \left\{ \overline{\Psi}^L \left(\frac{1}{g}M \! + \! U_{LR}\tilde{\Phi}_{LR}\right) \Psi^R \! + \! \overline{\Psi}^R\left(\frac{1}{g}M \! + \! U_{RL}\tilde{\Phi}_{RL} \right) \Psi^L \right\},\nonumber \\
 & &~  \label{action2}
\end{eqnarray}
from which we can derive the equation of motion for $\tilde{\Phi}_{LR}$ in the following form:
\begin{eqnarray}
 && \left[ \partial^2 + 4\bar{\partial}^2 \right]\left(\frac{1}{g}MU_{LR}+\tilde{\Phi}_{LR} \right)  \nonumber \\
 && +4gMU_{RL}\left\{\frac{1}{g}M \left( U_{LR}\tilde{\Phi}_{LR}+U_{RL}\tilde{\Phi}_{RL} \right) + \tilde{\Phi}_{LR}\tilde{\Phi}_{RL} \right\} \nonumber \\
 && +4g^2 \tilde{\Phi}_{LR}\left\{ \frac{1}{g}M \left( U_{LR}\tilde{\Phi}_{LR}+U_{RL}\tilde{\Phi}_{RL} \right) + \tilde{\Phi}_{LR}\tilde{\Phi}_{RL}\right\} \nonumber  \\
 && -g\kappa\overline{\Psi}^RU_{RL}\Psi^L =0.
\end{eqnarray}
If we here assume that $M^2 \propto \bar{x}^2$, then the term $4M^2\tilde{\Phi}_{LR}$ in the second line of the above equation yields the mass term for $\tilde{\Phi}_{LR}$, which should be compared with the bi-local field equation (1) in the COQM model. Further, for large $g\kappa$, $\overline{\Psi}^RU_{RL}\Psi^L$ becomes the leading term that determines $\tilde{\Phi}_{LR}$.

\section{Extension to a QCD-type Model }

The extension of the model in the previous section to a QCD-type model is not difficult, and is simply a formal task. In this case, the matter fields, the quark fields, have the components $\Psi^{L/R}_{i,\alpha}$, where $i~(=1,2,3)$ and $\alpha~(=1,2)$ are the indices of color and flavor, respectively; for simplicity, we confine our attention to the case of one generation. The Higgs-like scalar fields ${\mbf \phi}_{LR}$ are, then, $2\times 2$ matrices in flavor space and $8\times 8$ matrices in color space. We assume that the gauge degrees of freedom of those fields are factorized in such a way that ${\mbf \phi}_{LR}=U_L\tilde{\mbf \Phi}_{LR}U_R^\dagger$.

Now, writing the matter fields as ${\mbf \Psi}=\left[\matrix{{\mbf \Psi}^L \cr {\mbf \Psi}^R}\right]$, let us define the $\eta$-derivative of ${\mbf \Psi}$ by $\partial_\eta{\mbf \Psi}=i\pi{\mbf \Psi}=-i\pmatrix{0 & M_{LR} \cr M_{RL} & 0}{\mbf \Psi}$, where $M_{LR}=U_L M U_R^\dagger =U_{LR}M$, and $M$ is a function of $\bar{x}^2$. The meaning of $\eta$ is determined by the definition of $\partial_\eta{\mbf \Psi}$ itself. With the aid of this form of $i\pi$, it is not difficult to verify the nilpotency of the $\eta$-derivative operator (\ref{eta-derivative}) acting on matrices. The choice of $i\pi$ should be understood as a modified version of case (ii) in Eq. (\ref{vacuum_2}). 

According to the prescription in the previous  section, first, we define the covariant derivative operators ${\mbf D}^{L/R \pm}_\mu$ in this model by
\begin{equation}
 {\mbf D}^{L\pm}_\mu = ( D^L_\mu \pm \partial^R_\mu )\otimes T^L,~~~
 {\mbf D}^{R\pm}_\mu = ( D^R_\mu \pm \partial^L_\mu )\otimes T^R
\end{equation}
with
\begin{equation}
 D^{L/R}_\mu=\partial^{L/R}_\mu-ig{\mbf W}^{L/R}_\mu,~~~( {\mbf W}^{L/R}_\mu=W^a_\mu (x^{L/R}) T_a),
\end{equation}
where $T_a~(a=1 - 8)$ and $T^{L/R}$ are generators of color gauge transformations, normalized so that ${\rm tr}(T_aT_b)=\frac{1}{2}\delta_{a,b}$, and matrices defined in Eq.~(\ref{U(1/1)-generators}), respectively. There is no difference between the left and the right gauge fields, and hence, $L$ and $R$ simply designate their coordinate dependence. Secondly, the covariant derivatives ${\mbf D}_\eta$ and $\tilde{\mbf D}_\eta$ are again defined by
\begin{equation}
 {\mbf D}_\eta=\partial_\eta-ig\pmatrix{0 & {\mbf \phi}_{LR} \cr {\mbf \phi}_{RL} & 0 },~~~\tilde{\mbf D}_\eta=-ig\pmatrix{0 & {\mbf \Phi}_{LR} \cr {\mbf \Phi}_{RL} & 0},
\end{equation}
in the same way as in Eqs.~(\ref{bi-covariant_3}) and (\ref{bi-covariant_4}), except that ${\mbf \phi}_{LR}$ and ${\mbf \Phi}_{LR}=U_LMU_R^\dagger + \phi_{LR}$ are matrix fields obeying bilinear transformations such as ${\mbf \Phi}_{LR} \rightarrow {\cal U}_L{\mbf \Phi}_{LR}{\cal U}_R^\dagger$ under gauge transformations.

Using these covariant-derivative operators, the field strengths ${\mbf F}^{L/R\pm}_{\mu\nu}$, ${\mbf F}^{\pm}_{\mu\eta}$ and ${\mbf F}_{\eta\eta}$ are calculated similarly as those in the previous section, and we have
\begin{eqnarray}
 {\mbf F}^{L/R}_{\mu\nu} &=& {\mbf F}_{\mu\nu}(x^{L/R})\otimes T^{L/R},~({\mbf F}_{\mu\nu}=\frac{i}{g}[D_\mu,D_\nu]) , \\
 {\mbf F}_{\mu\eta}^{+} &=& ({\cal D}_\mu{\mbf \Phi}_{LR})Q + ({\cal D}_\mu{\mbf \Phi}_{RL})Q^\dagger ,  \nonumber \\
 {\mbf F}_{\mu\eta}^{-} &=& 2\left\{ (\bar{\cal D}_\mu{\mbf \Phi}_{LR})Q + (\bar{\cal D}_\mu{\mbf \Phi}_{RL})Q^\dagger \right\}
\end{eqnarray}
and
\begin{eqnarray}
{\mbf F}_{\eta\eta} &=& -2ig\left\{({\mbf \Phi}_{LR}{\mbf \Phi}_{RL}-\frac{1}{g^2}M^2)\otimes T^L+({\mbf \Phi}_{RL}{\mbf \Phi}_{LR}-\frac{1}{g}M^2)\otimes T^R \right\}.
\end{eqnarray}
Here,
\begin{eqnarray}
 {\cal D}_\mu {\mbf \Phi}_{LR} &=& \partial_\mu{\mbf \Phi}_{LR}-ig({\mbf W}_\mu^L{\mbf \Phi}_{LR}-{\mbf \Phi}_{LR}{\mbf}{\mbf W}_\mu^R), \\
 \bar{\cal D}_\mu {\mbf \Phi}_{LR} &=& \bar{\partial}_\mu{\mbf \Phi}_{LR}-\frac{1}{2}ig({\mbf W}_\mu^L{\mbf \Phi}_{LR}+{\mbf \Phi}_{LR}{\mbf W}_\mu^R), 
\end{eqnarray}
and the equations obtained by interchanging $R$ and $L$ also hold. The action for gauge and Higgs-like bi-local fields is given, again, by Eq. (\ref{action GH}); there, we do not need the second term on the right-hand side, since the ${\mbf F}^{L/R}_{\mu\nu}$ represent values at different spacetime points of one field. It is also apparent that the form of the action for the matter fields coincides with that of Eq.~(\ref{action M}). Therefore, the action in this QCD-type model should be given in the following form:
\begin{eqnarray}
 S &=& -\frac{1}{4}\int d^4x\,{\rm tr}\,{\mbf F}_{\mu\nu}{\mbf F}^{\mu\nu } \nonumber \\
 & & + \int d^4x^L\int d^4x^R {\rm Tr} \left\{ {\cal D}_\mu {\mbf \Phi}_{LR}{\cal D}_\mu {\mbf \Phi}_{RL} + 4\bar{\cal D}_\mu {\mbf \Phi}_{LR} \bar{\cal D}_\mu {\mbf \Phi}_{RL} \right\} \nonumber \\
 & & -2g^2 \int d^4x^L\int d^4x^R {\rm Tr} \left({\mbf \Phi}_{LR}{\mbf \Phi}_{RL}-\frac{1}{g^2}M^2 \right)^2 \nonumber \\
 & & +\int d^4x\overline{\mbf \Psi}i\gamma^\mu\left( \partial_\mu -ig{\mbf W}_\mu \right){\mbf \Psi} \nonumber \\
 & & + g\kappa \int d^4x^L\int d^4x^R \left(\overline{\mbf \Psi}^L{\mbf \Phi}_{LR}{\mbf \Psi}^R+\overline{\mbf \Psi}^R{\mbf \Phi}_{RL}{\mbf \Psi}^L \right) ,  \label{QCD action}
\end{eqnarray}
where tr and Tr stand for the traces taken over color indices and over color-flavor indices, respectively.

The next task is to choose a particular vacuum such that $\langle {\mbf \Phi}_{LR}{\mbf \Phi}_{RL} \rangle_0=\frac{1}{g^2}M^2$. As an extension of case (ii) in the previous section, we set
\begin{equation}
 {\mbf \Phi}_{LR} = U_L\left\{ \frac{1}{g}M+(\sigma_{LR}+i{\mbf \pi}_{LR})\right\}U_R^\dagger ,~~~({\mbf \pi}_{LR}=\pi^a_{LR}\tau_a ),
\end{equation}
where $\sigma_{LR}=\sigma(x^L,x^R)$ and $\pi^a_{LR}=\pi^a(x^L,x^R)$ are real fields.  Then, we have
\begin{eqnarray}
{\cal D}_\mu {\mbf \Phi}_{LR} &=& U_L \! \left[\partial_\mu \tilde{\mbf \Phi}_{LR} - ig({\mbf W}^{\prime L}_\mu - {\mbf W}^{\prime R}_\mu)\left\{\frac{1}{g}M+\tilde{\mbf \Phi}_{LR}\right\} \right] \! U^\dagger_R ,  \\
\bar{\cal D}_\mu {\mbf \Phi}_{LR} &=& U_L \! \left[\bar{\partial}_\mu \left\{ \frac{1}{g}M + \tilde{\mbf \Phi}_{LR} \right\} - \frac{i}{2}g({\mbf W}^{\prime L}_\mu + {\mbf W}^{\prime R}_\mu)\left\{\frac{1}{g}M+\tilde{\mbf \Phi}_{LR}\right\}  \right] \! U^\dagger_R ,~~~~~
\end{eqnarray}
where $\tilde{\mbf \Phi}_{LR}=\sigma_{LR}+i{\mbf \pi}_{LR}$ and
\begin{equation}
 {\mbf W}_\mu^{\prime K}=U_K^\dagger {\mbf W}^{K}_\mu U_K +\frac{i}{g}U^\dagger_{K}\partial_\mu U_{K},~~~(K=L,R). \label{W prime}
\end{equation}
Further, we have
\begin{equation}
 {\mbf \Phi}_{LR}{\mbf \Phi_{RL}} - \frac{1}{g^2}M^2 = U_L \left\{ \frac{1}{g}\left( M\tilde{\mbf \Phi}_{LR}+\tilde{\mbf \Phi}_{RL}M\right)+\tilde{\mbf \Phi}_{LR}\tilde{\mbf \Phi}_{RL}\right\}U_L^\dagger .
\end{equation}
In what follows, we do not distinguish ${\mbf W}^{\prime L/R}_\mu$ and ${\mbf W}^{L/R}_\mu$, since the kinetic terms of gauge fields are invariant under the transformation (\ref{W prime}). Therefore, recalling that ${\mbf \Phi}_{LR}={\mbf \Phi}(x^L,x^R)$ and ${\mbf \Phi}_{RL}={\mbf \Phi}^\dagger(x^R,x^L)$, the action (\ref{QCD action}) becomes

\begin{eqnarray}
 S &=& -\frac{1}{4}\int d^4x \, {\rm tr} \left({\mbf F}_{\mu\nu}{\mbf F}^{\mu\nu}\right)  \nonumber \\
 & & + \int d^4x^L\int d^4x^R \, {\rm Tr} \left\{ \left|{\partial}_\mu \sigma_{LR}-ig({\mbf W}^L_\mu - {\mbf W}^R_\mu)\left\{ \frac{1}{g}M+\sigma_{LR} \right\} \right|^2 \right.  \nonumber \\
 & &  + \, 4 \left| \bar{\partial}_\mu \left\{ \frac{1}{g}M + \sigma_{LR}\right\} - \frac{i}{2}g \left({\mbf W}^L_\mu+{\mbf W}^R_\mu \right) \left\{ \frac{1}{g}M+\sigma_{LR} \right\} \right|^2  \nonumber \\
 & & + \left| \partial_\mu {\mbf \pi}_{LR}-ig({\mbf W}^L_\mu - {\mbf W}^R_\mu){\mbf \pi}_{LR} \right|^2  \nonumber \\
 & & + \, 4 \left. \left| \bar{\partial}_\mu {\mbf \pi}_{LR}-\frac{i}{2}g({\mbf W}^L_\mu + {\mbf W}^R_\mu){\mbf \pi}_{LR} \right|^2 \right\} \nonumber \\
 & & -2g^2 \int d^4x^L\int d^4x^R \, Tr \left\{ \frac{1}{g}M(\sigma_{LR}+\sigma_{RL}) + \sigma_{LR}\sigma_{RL}+\vec{\pi}_{LR}\cdot\vec{\pi}_{RL}  \right. \nonumber \\
 & & \left. + \frac{i}{g}M({\mbf \pi}_{LR}-{\mbf \pi}_{RL}) + i({\mbf \pi}_{LR}\sigma_{RL}-\sigma_{LR}{\mbf \pi}_{RL})+i(\vec{\pi}_{LR}\times \vec{\pi}_{RL})\cdot \vec{\tau}  \right\}^2 \nonumber \\
 & & +\int d^4x\overline{\mbf \Psi}i\gamma^\mu ( \partial_\mu -ig{\mbf W}_\mu ){\mbf \Psi} \nonumber \\
 & & + g\kappa \int d^4x^L\int d^4x^R \left\{ \overline{\mbf \Psi}^L U_{LR}\left(\frac{M}{g}+\sigma_{LR}+i{\mbf \pi}_{LR} \right){\mbf \Psi}^R \right. \nonumber \\
 & & \left. +\overline{\mbf \Psi}^R U_{RL}\left( \frac{M}{g}+\sigma _{RL}-i{\mbf \pi}_{RL} \right){\mbf \Psi}^L \right\}. \label{QCD action2}
\end{eqnarray}

In the action (\ref{QCD action2}), neither the local color gauge symmetry nor the chiral flavor-$SU(2)$ symmetry remains, because of its non-local property and the presence of $M$. In particular, the mass terms for $\pi^a_{LR}$ fields arise only for $\pi^a(x^L,x^R)\neq \pi^a(x^R,x^L)$. The form of the resultant action has a similarity to that of the linear sigma model\footnote{Recently, the reality of $\sigma$ particle was discussed by S.~Ishida.\cite{sigma}} in an extended sense;  that is, if we take the local limit $x^L \rightarrow x^R$, then the $\pi^a_{LR}$ will tend to massless Goldstone, $\pi$ meson, fields. One can see that a non-trivial potential term $M^2(\bar{x}) \neq 0$ for $\sigma_{LR},\pi^a_{LR}$ arises simultaneously with the appearance of potential terms for ${\mbf \Psi}$ and ${\mbf W}^{L/R}_\mu$. We also note that if we regard $U_{LR}$ as Wilson's string function, then ${\mbf W}^L_\mu \pm {\mbf W}^R_\mu$ terms in (\ref{QCD action2}) vanish; and the action will restore the color gauge invariance in appearance.

\section{Summary and discussion}

In this paper, we have discussed the possibility of extending gauge theories based on NCG to a gauge theory including bi-local Higgs-like fields. The bi-local field theories following this line are also interesting from the point of view that they are low energy examples of matrix field theories.\cite{D-brane}

In \S 2, first, we reviewed Coquereaux's formulation of the NCG within the framework of the $U(1/1)$ gauge-Higgs model, in which the mass matrix of the matter fields, the Dirac fields, plays the role of the matrix coordinate connecting the left/right- components of matter fields at the same spacetime point. The mass parameter in the matrix coordinate determines the order of symmetry breaking. In addition, the Higgs-like fields can be understood as the quantum fluctuations around this matrix coordinate. Then, we attempted to extend such a gauge-Higgs theory to a bi-local theory by introducing a matrix coordinate that connects the left/right-components of matter-fields at different spacetime points. With this extension, the Higgs-like fields in this model become bi-local fields, while the gauge fields and the matter fields remain as local fields living in the left or in the right world. Furthermore, the mass parameter in the matrix coordinate is allowed to be a function of relative coordinates. Then it acquires the meaning of a c-number potential function for the bi-local Higgs-like fields. 

The extension of the toy model to a QCD-type model is rather formal, and this was done in \S 3 within the framework of one generation. The matter fields in this case are assumed to be $(u,d)$-quark fields, and the gauge structure in this model is the color gauge symmetry, which does not distinguish between the left and the right gauge fields. In this extended model, the mass parameter is introduced through the $\eta$-derivative of matter fields: $(\partial_\eta)_{LR}=-iU_LMU_R^\dagger$, etc. Here, $U_L$ and $U_R$ are necessary to guarantee the gauge invariance of the theory before symmetry breaking. In particular, if we regard $U_L$ and $U_R$ as Wilson string functions, then the $\overline{\mbf \Psi}_LU_{LR}{\mbf \Psi}_R$ can be leading terms determining bi-local fields under a specific choice of parameters.

 The extended bi-local fields in this QCD-type model should be regarded as mesons, although its action contains non-local 3-body and 4-body self-interactions. The resultant action has a structure similar to that of the linear sigma model in local-chiral field theories; the mass terms, potential terms, for $\pi$-meson components vanish in the local limit $x^L \rightarrow x^R$. It should be noted that if we replace $M$ with a matrix in flavor space, such as ${\mbf M}=a{\mbf 1}+b\tau_3$, in the sense of explicit symmetry breaking, then the $\pi^\pm$-components of the bi-local field remain massive fields even in the local limit. It is also important that if we assume a functional form such that $M^2 \propto \bar{x}^2$, the model has a structure similar to COQM. Unfortunately, however, there is no way to determine the functional form of $M^2$ from this formalism. Under such an assumption, further, the elimination of the relative time $\bar{x}^0$ becomes a serious problem, as in COQM. Also, in our approach of introducing bi-local fields ${\mbf \Phi}_{LR}$, the role of the intrinsic spins in quark-bound states is not clear.

Although these problems are left as the subject of future examinations, the bi-local extension of gauge theories based on non-commutative geometry gives us an interesting insight to study the bound-state structure of matter fields. \\

\begin{center}
{\bf Acknowledgements}
\end{center} \vspace{5mm}

The authors would like to express their sincere thanks to Professor S.~Ishida and Professor S.~Y.~Tsai for their interest and discussions. One of the authors (S.~N.) wishes to express his gratitude to the late Dr. H.~Suura for his discussions in the early stage of this work. They are also grateful to Professor J.~Otokozawa, Dr. S.~Deguchi and other members of their laboratory for encouragement.


\begin{thebibliography}{99}
%
\bibitem{meson}
For example, A.~A.~Migdal, Ann.~of Phys.\ {\bf 126} (1980), 279.
\\
W.~Lucha, F.~Sch\"oberl, D.~Gromes, Phys.~Rep.\ {\bf 200} (1991), 127.
\\
See also references contained therein.
\bibitem{Yukawa}
H.~Yukawa, Phys.~Rev.\ {\bf 91} (1953), 415, 416.
\bibitem{bi-local}
As for the review articles of bi-local models, see the following:
\\
T.~Takabayasi, Prog.~Theor.~Phys.~Suppl.\ vol.~67 (1979), 1.
\\
T.~Got\=o, S.~Naka and K.~Kamimura, ibid, 69.
\\
T.~Takabayasi, Nuovo Cim. {\bf 33} (1964), 668.
\bibitem{COQM}
S.~Ishida and T.~Sonoda, Prog.~Theor.~Phys.\ {\bf 70} (1984), 1323.
\\
S.~Ishida and M.~Oda, in {\it Proc. of the International Symposium on Extended Objects and Bound Systems, Karuizawa, Japan, 19-21 March 1992}, ed. O.~Hara et al. (World Scientific Pub Co., 1992), p181.
\bibitem{Connes}
A.~Connes and J.~Lott, Nucl.~Phys.\ {\bf B} (Proc.~Suppl.) {\bf 18} (1990), 29.
\\
See also A.~Connes, {\it Noncommutative Geometry} (Academic Press, New York, 1994).
\bibitem{many authors}
A.~H.~Chamseddine, G.~Felder and J.~Fr\"ohlich, Phys.~Lett.\ {\bf B296} (1992), 109; Nucl.~Phys.\ {\bf B395} (1993), 672.
\\
D.~Kastler, Rev.~Math.~Phys.\ {\bf 5} (1993), 477.
\\
A.~Sitarz, Phys.~Lett.\ {\bf B308} (1993), 311.
\\
K.~Morita, Prog.~Theor.~Phys.\ {\bf 90} (1993), 219.
\\
H.~G.~Ding, H.~Y.~Guo, J.~M.~Li and K.~Wu, Z.~Phys.\ {\bf C64} (1994), 521.
\\
K.~Morita and Y.~Okumura, Prog.~Theor.~Phys.\ {\bf 91} (1994), 959.
\\
Y.~Okumura, Prog.~Theor.~Phys.\ {\bf 92} (1994), 625.
\\
G.~Konisi and T.~Saito, Prog.~Theor.~Phys.\ {\bf 95} (1996), 657.
\\
See also J.~Madore, Phys.~Rev.\ {\bf D41} (1990), 3709.
\bibitem{early attempt}
S. Naka, E. Umezawa, T. Matsufuji, S. Abe and Y. Furukawa, {\it Proc. of the Workshop on Fundamental Problem in Particle Physics}, NUP-A-95-11.
\bibitem{matrix formulation}
R.~Coquereaux, G.~Esposito-Farese and G.~Vaillant, Nucl.~Phys.\ {\bf B353} (1991), 689.
\\
See also S.~Naka and E.~Umezawa, Prog.~Theor.~Phys.\ {\bf 92} (1994), 189.
\bibitem{sigma}
S.~Ishida, M.~Y.~Ishida, T.~Ishida, K.~Takamatsu and T.~Tsuru, Prog.~Theor.~Phys.\ {\bf 98} (1997), 621.
\bibitem{D-brane}
P.-M.~Ho and Y.-S.~Wu, Phys.~Lett.\ {\bf B398} (1997), 52.
\\
M.~R.~Douglas, hep-th/9901146.
\\
D.~Bak and S.-J.~Rey, hep-th/9902101.
\\
K.~Hashimoto, hep-th/9903115.
\\
See also E.~Witten, Nucl.~Phys.\ {\bf B460} (1996), 335.
\\
Z.~Kakushadze and S.-H.~Henry~Tye, hep-th/9809147.
\end{thebibliography}
\end{document}